\documentclass[12pt]{iopart}
\usepackage{iopams}  
\begin{document}

\title{Two oscillators in a common heat bath}

\author{R. F. O'Connell}

\address{Department of Physics and Astronomy \\
Louisiana State University, Baton Rouge, LA 70803-4001 USA}
\ead{oconnell@phys.lsu.edu}
\begin{abstract}
We show that the case of two oscillators in a common heat bath cannot be reduced to an effective one body problem.  In addition, there is an interaction between the oscillators, even at zero temperature, due to the fluctuations caused in both oscillators by the zero-point oscillations of the electromagnetic field.\end{abstract}

\maketitle

The case of a quantum particle in a general heat bath at an arbitrary temperature $T$ is now generally well understood \cite{ford85,ford88}.  The key starting point is the independent oscillator (IO) Hamiltonian, which has a lower bound \cite{ford88}.

Quantum information scenarios now underline the necessity of considering many body systems.  The simplest system of this kind to analyze is the case of 2 particles in a common heat bath, which was considered in \cite{paz09} and \cite{chou08}.  The approach used by these authors was to start with a two-body Hamiltonian, after which they made a transformation to center-of-mass and relative coordinates.  Their general conclusion was that the two harmonic oscillator model could be separated such that the relevant coordinate motion has no interaction with either the center-of-mass motion or with the oscillators of the heat bath.  However, both \cite{paz09} and \cite{chou08} have a major omission, the result of which is that their Hamiltonian has no lower bound.  Thus, we are motivated to consider the same problem but using the exact Hamiltonian which ensures a lower bound.  The result of our analysis is that it is not possible to reduce the 2 particle problem to a 1 particle problem, as we will now prove.

First, we recall the exact one body IO quantum Hamiltonian \cite{ford88}

\begin{equation}
H_{IO}=\frac{p^{2}}{2m}+V(x)+\sum_{j}\left(\frac{p^{2}_{j}}{2m_{j}}+m_{j}\omega^{2}_{j}\left(q_{j}-x\right)^{2}\right). \label{chb1}
\end{equation}
We emphasize, in particular the factor $(q_{j}-x)^{2}$ in the interaction term, which ensures that $H$ has a lower bound \cite{ford88}. Next, using the Heisenberg equations of motion leads to a quantum Langevin equation for a quantum particle of mass $m$ moving in an oscillator potential in an arbitrary heat bath and temperature $T$:
\begin{equation}
m\ddot{x}+\int^{t}_{-\infty}dt_{1}\mu\left(t-t_{1}\right)\dot{x}\left(t_{1}\right)+m\omega^{2}x=F(t), \label{chb2}
\end{equation}
where the dot and prime denote, respectively, the derivative with respect to $t$ and $x$.  This is a Heisenberg equation of motion for the coordinate operator $x$.  The coupling with the heat bath is described by two terms: an operator-valued random force $F(t)$ with mean zero, and a mean force characterized by the memory function $\mu(t)$.  These quantities are given in terms of the heat bath variables:
\begin{equation}
\mu (t)=\sum_{j}m_{j}\omega^{2}_{j}\cos (\omega_{j}t)\Theta(t), \label{chb3}
\end{equation}
where $\Theta (t)$ is the Heaviside step function (by convention the memory function vanishes for negative times), and
\begin{equation}
F(t)=\sum_{j}m_{j}\omega^{2}_{j}q^{h}_{j}(t), \label{chb4}
\end{equation}
were $q^{h}_{j}(t)$ denotes the general solution of the homogeneous equation for the heat bath oscillators (corresponding to no interaction).  We also note that, in the particular case of Ohmic dissipation, equation (\ref{chb2}) reduces to
\begin{equation}
m\ddot{x}+m\gamma\dot{x}+m\omega^{2}x=F(t). \label{chb5}
\end{equation}
The generalization to 2 bodies is straightforward so that our two body Hamiltonian (taking the masses and spring constants of both oscillators to be the same, as in \cite{chou08}) may be written as
\begin{eqnarray}
H &=& \frac{p^{2}_{1}}{2m}+ \frac{1}{2}m\omega^{2}x^{2}_{1}+\sum_{j}\left[\frac{p^{2}_{j}}{2m_{j}}+\frac{1}{2}m_{j}\omega^{2}_{j}\left(q_{j}-x_{1}\right)^{2}\right] \nonumber \\
&+& \frac{p^{2}_{2}}{2m}+\frac{1}{2}m\omega^{2}x^{2}_{2}+\sum_{j}\left[\frac{p^{2}_{j}}{2m_{j}}+\frac{1}{2}m_{j}\omega^{2}_{j}\left(q_{j}-x_{2}\right)^{2}\right]  \nonumber \\
&-& \left[\frac{p^{2}_{j}}{2m_{j}}+\frac{1}{2}m_{j}\omega^{2}_{j}q^{2}_{j}\right] \nonumber \\
&=& \frac{1}{2m} \left(p^{2}_{1}+p^{2}_{2}\right)+\frac{m}{2}\omega^{2}\left(x^{2}_{1}+x^{2}_{2}\right) \nonumber \\
&& ~+\sum_{j}\left[\frac{p^{2}_{j}}{2m_{j}}+\frac{1}{2}m_{j}\omega^{2}_{j}q^{2}_{j}\right] \nonumber \\
&&~ -\sum_{j}m_{j}\omega^{2}_{j}q_{j}(x_{1}+x_{2}) \nonumber \\
&&~ + \sum_{j}\frac{1}{2}m_{j}\omega^{2}_{j}\left(x^{2}_{1}+x^{2}_{2}\right). \label{chb6}
\end{eqnarray}
We emphasize that our Hamiltonian contains a lower bound, manifest by the square terms in the interaction of $x_{1}$ and $x_{2}$ with the heat bath $(q_{j})$.  We also note that it has the correct non-interacting part.

Next, we transform to center-of-mass and relative coordinates
\begin{equation}
x=x_{1}-x_{2};~~~X=(x_{1}+x_{2})/2, \label{chb7}
\end{equation}
\begin{equation}
m_{x}=m/2;~~~M=2m, \label{chb8}
\end{equation}
and
\begin{equation}
p=\frac{1}{2}(p_{1}-p_{2});~~~P=p_{1}+p_{2}. \label{chb9}
\end{equation}
As a result, (\ref{chb6}) may be written as
\begin{equation}
H=H_{cm}+H_{rel} \label{chb10}
\end{equation}
where
\begin{eqnarray}
H_{cm} &=& \frac{P^{2}}{2M}+\frac{1}{2}M\omega^{2}X^{2}+2\sum_{j}\frac{p^{2}_{j}}{2m_{j}}+2\sum_{j}\frac{1}{2}m_{j}\omega^{2}_{j}q_{j}^{2} \nonumber \\
&-& \sum_{j}2m_{j}\omega^{2}_{j}q_{j}X+\sum m_{j}\omega^{2}_{j}X^{2}, \label{chb11}
\end{eqnarray}
and
\begin{equation}
H_{rel}=\frac{p^{2}}{2m_{x}}+\frac{1}{2}m_{x}\omega^{2}x^{2}+\frac{1}{4}x^{2}\sum_{j}m_{j}\omega^{2}_{j}-\left[\frac{p^{2}_{j}}{2m_{j}}+\frac{1}{2}m_{j}\omega^{2}_{j}q^{2}_{j}\right]. \label{chb12}
\end{equation}
Thus, $H_{cm}$ may be re-written in the form
\begin{equation}
H_{cm} = \frac{P^{2}}{2M}+\frac{1}{2}M\omega^{2}X^{2}+2\sum_{j}\left[ \frac{p^{2}_{j}}{2m_{j}}+\frac{1}{2}m_{j}\omega^{2}_{j}\left(q_{j}-X\right)^{2}\right] .\label{chb13}
\end{equation}
Redefining the bath parameters as follows
\begin{equation}
m_{j}=2\tilde{m}_{j},~~\omega_{j}=\frac{\tilde{\omega}_{j}}{2} , \label{chb14}
\end{equation}
we obtain
\begin{equation}
H_{cm} = \frac{P^{2}}{2M}+\frac{1}{2}M\omega^{2}X^{2}+\sum_{j}\left[\frac{p^{2}_{j}}{2m_{j}}+\frac{1}{2}\tilde{m}_{j}\tilde{\omega}^{2}_{j}\left(q_{j}-X\right)^{2}\right] . \label{chb15}
\end{equation}
Thus, the centre-of-mass motion is reduced to a 1-body problem.  Also, since we are summing over $j$, it is clear that the Hamiltonian has the correct form of the one-body problem, given in (\ref{chb1}).  However, $H_{rel}$ cannot be written in this form.  Thus, this two-body problem cannot be reduced to an effective one-body problem.

Also, since the equation of motion of both $x_{1}$ and $x_{2}$ involve the parameters of the heat bath, it is clear that there is an interaction, via the heat bath, between $x_{1}$ and $x_{2}$. This is analogous to what occurs for the London and Casimir-Polder effects \cite{power64,london26,casimir48} as well as the Casimir effect \cite{milton01}.  For the present case of two oscillators in a heat bath, we are in the macroscopic domain so that the corresponding results are thus Casimir-like and need to be calculated, which we will address in the future.
Moreover, the equation of motion of oscillator 1 at time $t$ contains contributions from $F_{1}(t)$ and also from oscillator 2.  But since $x_{2}$ is also affected by $F_{1}(t)$, it is clear that its appearance in the equation of motion of $x_{1}$ must be at the earlier time $t-\tau$ to ensure that it is affected by $F_{1}(t)$ at time $t$.  The equation of motion of oscillator 2 is similarly affected.  The end result is that the equations of motion of $x_{1}$ and $x_{2}$ are interconnected and quite complicated so that a numerical analysis will be needed to obtain exact results.

\ack{This work was partially supported by the National Science Foundation under Grant No. ECCS-1125675.}


\newpage





\begin{thebibliography}{7}

\bibitem{ford85} G. W. Ford, J. T. Lewis, and R. F. O'Connell, "Quantum Oscillator in a Blackbody Radiation Field," \emph{Phys. Rev. Lett.} \textbf{55}, 2273 (1985).

\bibitem{ford88} G. W. Ford, J. T. Lewis, and R. F. O'Connell, "The Quantum Langevin Equation," \emph{Phys. Rev. A} \textbf{37}, 4419 (1988).

\bibitem{paz09} J. P. Paz and A. J. Roncaglia, \emph{Phys. Rev. Lett.} \textbf{100}, 220401 (2008); ibid. \emph{Phys. Rev. A} \textbf{79}, 032102 (2009).

\bibitem{chou08} C.-H. Chou, T. Yu, and B. L. Hu, \emph{Phys. Rev. E} \textbf{77}, 011112 (2008).

\bibitem{power64} E. A. Power, "Introductory Quantum Electrodynamics" (Longmans 1964).

\bibitem{london26} F. London, \emph{Zeits. fur Physik} \textbf{63}, 245 (1926).

\bibitem{casimir48} H. B. G. Casimir and D. Polder, \emph{Phys. Rev.} \textbf{73}, 360 (1948).

\bibitem{milton01} K. A. Milton, "The Casimir Effect" (World Scientific, 2001).













\end{thebibliography}
\end{document}